\documentclass[times,twocolumn,final]{elsarticle}
\usepackage{medima}
\usepackage{framed,multirow}
\usepackage{multirow}
\usepackage{amssymb}
\usepackage{latexsym}
\usepackage{hyperref}
\usepackage{amsmath}

\definecolor{newcolor}{rgb}{.8,.349,.1}
\hypersetup{
    colorlinks=true
}
\usepackage{url}
\usepackage{xcolor}
\usepackage{subfigure}

\usepackage{hyperref}

\definecolor{newcolor}{rgb}{.8,.349,.1}
\newcommand{\modified}[1]{\textcolor{black}{#1}}
\journal{Radiotherapy and Oncology}

\begin{document}

\verso{Yixing Huang \textit{et~al.}}

\begin{frontmatter}

%\title{A Multicenter Study on Deep Learning for Brain Metastases Detection: Data Heterogeneity and Data Privacy}
\title{{Multicenter Privacy-Preserving Model Training for Deep Learning Brain Metastases Autosegmentation}}
%\tnoteref{tnote1}}%
%\tnotetext[tnote1]{This is an example for title footnote coding.}

\author[1,4,5]{Yixing Huang\corref{cor1}}
\cortext[cor1]{Corresponding authors: yixing.yh.huang@fau.de; $\qquad$ stephanie.tanadini-lang@usz.ch}
\author[2]{Zahra Khodabakhshi}
\author[1,4,5]{Ahmed Gomaa}
\author[3]{Manuel Schmidt}
\author[1,4,5]{Rainer Fietkau}
\author[2]{Matthias Guckenberger}
%% Third author's email
%\ead{author3@author.com}
\author[2]{Nicolaus Andratschke}
\author[1,4,5]{Christoph Bert}
\author[2]{Stephanie Tanadini-Lang\corref{cor1}}
\author[1,4,5]{Florian Putz}

\address[1]{Department of Radiation Oncology, Universit\"atsklinikum Erlangen, Friedrich-Alexander-Universit\"at Erlangen-N\"urnberg, Erlangen, Germany }
\address[2]{Department of Radiation Oncology, University Hospital Zurich, University of Zurich, Zurich, Switzerland}
\address[3]{Department of Neuroradiology, Universit\"atsklinikum Erlangen, Friedrich-Alexander-Universit\"at Erlangen-N\"urnberg, Erlangen, Germany}
\address[4]{Comprehensive Cancer Center Erlangen-EMN (CCC ER-EMN),
Erlangen, Germany}
\address[5]{Bavarian Cancer Research Center (BZKF),
Erlangen, Germany}

%\received{1 May 2013}
%\finalform{10 May 2013}
%\accepted{13 May 2013}
%\availableonline{15 May 2013}
%\communicated{S. Sarkar}

\begin{abstract}
%%%
\textbf{Objectives:} 
This work aims to explore the impact of multicenter data heterogeneity on deep learning brain metastases (BM) autosegmentation performance, and assess the efficacy of an incremental transfer learning technique, namely learning without forgetting (LWF), to improve model generalizability without sharing raw data.

\textbf{Materials and methods:} 
A total of six BM datasets from University Hospital Erlangen (UKER), University Hospital Zurich (USZ), Stanford, UCSF, New York University (NYU), and BraTS Challenge 2023 were used. First, the performance of the DeepMedic network for BM autosegmentation was established for exclusive single-center training and mixed multicenter training, respectively. Subsequently privacy-preserving bilateral collaboration was evaluated, where a pretrained model is shared to another center for further training using transfer learning (TL) either with or without LWF.

\textbf{Results:} For single-center training, average F1 scores of BM detection range from 0.625 (NYU) to 0.876 (UKER) on respective single-center test data. Mixed multicenter training notably improves F1 scores at Stanford and NYU, with negligible improvement at other centers. When the UKER pretrained model is applied to USZ, LWF achieves a higher average F1 score (0.839) than naive TL (0.570) and single-center training (0.688) on combined UKER and USZ test data. Naive TL improves sensitivity and contouring accuracy, but compromises precision. Conversely, LWF demonstrates commendable sensitivity, precision and contouring accuracy. When applied to Stanford, similar performance was observed.

\textbf{Conclusion:}
Data heterogeneity {(e.g., variations in metastases density, spatial distribution, and image spatial resolution across centers)} results in varying performance in BM autosegmentation, posing challenges to model generalizability. LWF is a promising approach to peer-to-peer privacy-preserving model training.

%%%%
\end{abstract}

\begin{keyword}
%% MSC codes here, in the form: \MSC code \sep code
%% or \MSC[2008] code \sep code (2000 is the default)
%\MSC 41A05\sep 41A10\sep 65D05\sep 65D17
%% Keywords
\KWD Continual learning\sep Learning without forgetting\sep Brain metastasis\sep Multicenter collaboration\sep Data privacy\sep Data heterogeneity 
\end{keyword}

\end{frontmatter}

%\linenumbers

%% main text
\section{Introduction}
\label{sect:Intro}

Individuals diagnosed with metastatic cancer face a significant likelihood of developing brain metastases (BM), with reported incidence rates reaching up to 40\% \cite{tabouret2012recent}.
Stereotactic radiosurgery (SRS) has emerged as a preferred method for BM treatment \cite{rogers2022stereotactic}.  
The planning of SRS treatment necessitates detailed information on the number, size, locations, and boundaries of BM \cite{welzel2022stereotactic}, which in turn requires their precise detection and contouring. 
Manual BM identification is not only time-intensive but also prone to variability between observers \cite{xue2020deep}. This manual identification process particularly fails with smaller metastases, which may go unnoticed due to their visibility on few image slices and their low contrast \cite{xue2020deep}. Moreover, the challenge of distinguishing BM from similar-looking anatomical structures, like blood vessels, in 2D image slices further complicates manual identification \cite{kocher2020applications}. 
Consequently, the development of automated, computer-assisted techniques for BM identification holds significant importance for clinical practice.

%\cite{buchner2023development,yin2022development,ottesen20232,qu2023construction} \cite{fairchild2023deep,pennig2021automated,charron2018automatic,huang2022deep}
With the fast development of deep learning techniques and their adoptions into the field of radiation oncology, many deep learning algorithms have been proposed for automatic BM detection and segmentation \cite{wang2023brain,ozkara2023deep}, which have achieved high efficiency (compared with manual identification) and impressive efficacy despite of performance variance. According to the latest systematics reviews \cite{wang2023brain,ozkara2023deep}, 3D U-Net and DeepMedic based networks are the most commonly used and effective networks for BM identification. Among all the reported methods, most of them were developed from and evaluated on single-center, in-house data, while multi-center studies are gaining more and more attention \cite{xue2020deep,buchner2023development,qu2023construction,yin2022development,ottesen20232,liew2023gradual,bouget2022preoperative}. 
 The performance of deep learning algorithms relies highly on the amount and quality of training data. Due to the limited data in a single center, multicenter collaboration is very important for developing high performance deep learning models. Nevertheless, the influence of data heterogeneity among multiple centers in deep learning auto-segmentation model performance has not yet been fully addressed.
 
In addition, due to data privacy and data management regulations (e.g., the EU medical device regulation \citep{beckers2021eu}), data sharing among multiple centers is restricted, which impedes the development of high performance deep learning tumor segmentation models from multicenter collaboration.
To overcome the data privacy issue, federated learning, which trains a high performance model collaboratively without sharing data, has been proposed \cite{xu2021federated}. Due to the technical difficulty, communication frequency, financial cost and management complexity, center-to-peer federated learning \cite{xu2021federated}, where a central server is required to coordinate training information for a global model, is challenging for practical use. Moreover, an implicit power hierarchy may arise from the centralized structure of center-to-peer federated learning, making cooperation unattractive for participating peer-level institutions. Therefore, peer-to-peer federated learning \cite{xu2021federated} is more feasible in practice, and the simplest way is to continually train the same model one center after another \cite{sheller2020federated}. However, when a model is retrained on new datasets or tasks, deep learning suffers from the problem of catastrophic forgetting \citep{delange2021continual}, i.e., deep learning models forget learned old knowledge catastrophically. Continual learning \cite{delange2021continual} aims to allow machine learning models to be updated through new data while retaining previously learned knowledge. In our in-depth technical survey \cite{huang2023survey}, learning without forgetting (LWF) \cite{li2017learning} was superior to other regularization-based continual learning methods with statistical significance. Therefore, in this work, the efficacy of LWF in multicenter collaboration on BM autosegmentation is investigated. 

{This work aims to:
\begin{itemize}
\item Investigate the influence of data heterogeneity across multiple centers on the performance of deep learning models for BM autodetection.
\item Assess the efficacy of LWF as a privacy-preserving strategy for multicenter collaboration in improving model generalizability.
\item \modified{Evaluate a single-center training model to be used at other centers.}
\end{itemize}
}

\section{Materials And Methods}
\subsection{Datasets}

\begin{table*}[t]
\caption{Details of BM datasets from different centers.}
\label{Tab:Datasets}
\centering
\begin{tabular}{lcccccc}
\noalign{\hrule height 1.5pt}
 Dataset & UKER & USZ & Stanford & UCSF & NYU & BraTS\\
\noalign{\hrule height 1.5pt}
$\#$ training volumes &600 & 157 &56 &200 & 104 &178\\
\hline
$\#$ validation volumes &67 & 10 &5 &23 & 10 &10\\
\hline
$\#$ test volumes &103 & 35 &40 &100 & 50 & 50\\
\noalign{\hrule height 1.5pt}
{In-plane resolution (mm)} &$\leq$1.0 &0.6 &0.94 &$\leq$1.0 &$\leq$1.0 & $\leq$1.0\\
\hline
{Through-plane slice thickness (mm)} & $\leq$1.0 &0.6 &1.0 & $\leq$1.5&$\leq$5.0 &$\leq$5.0\\
\noalign{\hrule height 1.5pt}
$\#$ metastases per volume & 2.2 &4.2 &12.2 & 10.3&6.6 &7.8\\
\hline
$\#$ training metastases &1505 &640 &544 &2159 &616 &1601\\
\hline
$\#$ validation metastases &130 &29 &33 & 357&102 &35\\
\hline
$\#$ test metastases &272 &144 &656 &826 &359 &213\\
\noalign{\hrule height 1.5pt}
Mean training metastasis size (cm$^3$) &1.642 & 1.506&0.352 &0.327 &0.828 &1.371\\
\hline
Mean validation metastasis size (cm$^3$) &1.761 &2.650 &0.196 &0.329 &0.583 &5.603\\
\hline
Mean test metastasis size (cm$^3$) &1.950 &1.490 &0.237 &0.879 &1.189 &1.786\\
\noalign{\hrule height 1.5pt}
Median training metastasis size (cm$^3$) &0.132 & 0.229&0.054 & 0.034& 0.078&0.073\\
\hline
Median validation metastasis size (cm$^3$) &0.216 & 0.402&0.067 & 0.038& 0.077&0.070\\
\hline
Median test metastasis size (cm$^3$) &0.131 & 0.134 &0.031 &0.037 &0.077 &0.067\\
\noalign{\hrule height 1.5pt}
Training metastases $\leq$ 0.1 cm$^3$ & 44.4\% &35.4\% &67.5\% &71.2\% &55.5\% &56.3\%\\
\hline
Validation metastases $\leq$ 0.1 cm$^3$ & 36.9\% &31.0\% &60.6\% &72.3\% &57.8\% &54.3\%\\
\hline
Test metastases $\leq$ 0.1 cm$^3$ & 46.0\% &44.4\% &79.0\% &67.4\% &57.4\% &55.9\%\\
\noalign{\hrule height 1.5pt}
\end{tabular}
\end{table*}

T1 contrast enhanced (T1CE) MRI datasets from University Hospital Erlangen (Universit\"atsklinikum, UKER), University Hospital Zurich (Universit\"ats Spital Z\"urich, USZ), Stanford University \cite{grovik2020deep}, University of California San Francisco (UCSF) \cite{rudie2023university}, New York University (NYU) ({\url{https://nyumets.org/}) and Brain Tumor Segmentation (BraTS) 2023 BM Segmentation Challenge \cite{moawad2023brain} were used. \modified{The UKER and USZ datasets are internal, with no ethical review needed for this study, as patients provided written consent for retrospective scientific use of their data. The other datasets are publicly available with their respective Institutional Review Board (IRB) approval and Data Transfer Agreement (DTA) \cite{grovik2020deep,rudie2023university,moawad2023brain}.} Since T1CE is the main sequence used for radiotherapy treatment planning \cite{putz2024quality,kaufmann2020consensus,buchner2023identifying}, in this work only T1CE volumes are used.

The UKER dataset contains 853 {T1CE} volumes in total using the MPRAGE sequence from a longitudinal study \citep{putz2020fsrt}. The volumes were acquired from various Siemens Healthcare 1.5 Tesla MRI scanners (Magnetom Aera and Magnetom Avanto mainly). The primary cancers include 41.3\% skin cancer, 22.2\% lung cancer, 12.4\% breast cancer, and 10.5\% kidney cancer. 
The USZ dataset contains 202 {T1CE} turbo field echo (TFE) volumes from a Philips Healthcare Ingenia 3.0 Tesla scanner, with lung and melanoma primary cancers. 
The details of other public datasets can be found in their respective descriptions \cite{grovik2020deep,rudie2023university,moawad2023brain}. 
\modified{Note that the total numbers of cases with labels accessible to public participants may differ from those reported in the dataset descriptions, since certain test cases are internal only. The provided labels were used without further refinement to avoid potential biases.} The BM distribution of different datasets in size, density, and patient number are summarized in Tab.\,\ref{Tab:Datasets}. Each dataset is exclusively partitioned into training, validation, and test subsets, with all volumes from the same patient contained within a single subset. All the MRI volumes were preprocessed by the same pipeline:  skull stripping, volume/voxel size uniformization, bias field correction, and intensity Z-normalization. All the volumes have $240 \times 240 \times 155$ voxels with a voxel size of $1\,\textrm{mm} \times 1\,\textrm{mm} \times 1\,\textrm{mm}$.

\subsection{Neural network and learning without forgetting}
The DeepMedic network \cite{kamnitsas2017efficient} is chosen because of its efficacy in various brain tumor segmentation tasks as well as its success in BM identification \citep{kamnitsas2017efficient,liu2017deep,lu2019automated,charron2018automatic,hu2019multimodal,huang2022deep}.  
In our previous work \cite{huang2022deep}, DeepMedic has achieved encouraging performance on the UKER dataset. In this work, the performance of DeepMedic on different datasets is evaluated.

In order to train a model from multiple centers without sharing raw data, peer-to-peer federated learning is one promising approach. {The most straightforward way for such peer-to-peer federated learning involves continuous model training across various centers through weight transfer. Specifically, single weight transfer (SWT) sequentially trains the model from the first to the last center, whereas cyclic weight transfer (CWT) iteratively trains the model across all centers in cycles, as described in \cite{chang2018distributed}.} To avoid the forgetting problem after model sharing, continual learning techniques can be applied. In our previous in-depth survey \cite{huang2023survey}, LWF was demonstrated to have superior performance to other continual learning regularization methods. Therefore, LWF is chosen in this work. LWF controls forgetting by imposing network output stability. In other words, the model trained with local data should yield similar predictions for specific samples as it did before training. Knowledge distillation loss (KDL) \cite{hinton2015distilling} is the key element of LWF, which constrains the new model (called student model) output to have a similar distribution of class probabilities to those predicted by the teacher model.

The objective function for LWF consists of a regular segmentation loss $\mathcal{L}_{\textrm{seg}}$ and the KDL $\mathcal{L}_{\textrm{KDL}}$,
\begin{equation}
\begin{array}{l}
\mathcal{L}_{\text{LWF}} = \mathcal{L}_{\textrm{seg}}\left(\mathcal{M}(\boldsymbol{x}, \boldsymbol{\theta}), \boldsymbol{y}\right) + 
\lambda \mathcal{L}_{\textrm{KDL}}\left(\mathcal{M}(\boldsymbol{x}, \boldsymbol{\theta}),  \mathcal{M}\left(\boldsymbol{x}, \boldsymbol{\theta}_0\right)\right),
\end{array}
\label{eq:LwFLoss}
\end{equation}
where $\mathcal{M}$ is the network model, $\boldsymbol{x}$ is the set of input data samples and $\boldsymbol{y}$ is the set of corresponding ground truth segmentation masks, $\lambda$ is a relaxation parameter for KDL, $\boldsymbol{\theta}_0$ is the model parameter set from the previous center (teacher model), and $\boldsymbol{\theta}$ is the current model parameter set to optimize.

\subsection{\modified{Experimental Setup}}

All DeepMedic models were trained on an NVIDIA Quadro RTX 8000 GPU with Intel Xeon Gold 6158R CPUs. The model was trained for 300 epochs at each center. The validation was performed every two epochs, and the final models were selected based on the best validation performance (i.e., best volumetric Dice scores). The Adam optimizer with an initial learning rate of 0.001 and a weight decay of 0.0001 was applied. A probability of 50\% for extracting tumor class containing segments was applied to keep class-balance. In this work, the binary cross-entropy (BCE) loss together with a subvolume-level sensitivity-specificity loss ($\alpha = 0.5$) \cite{huang2022deep} was used for $\mathcal{L}_{\textrm{seg}}$, and $\lambda$ was set to 0.1 in Eqn.\,(\ref{eq:LwFLoss}). Our implementation is publicly available on GitHub for further research and collaboration\footnote{\url{https://github.com/YixingHuang/DeepMedicPytorch}}.

\modified{In order to test the applicability of a single-center trained model for other centers, the UKER pretrained (our own) model was shared to USZ and Stanford respectively, where a bilateral agreement was available.} 

For detection performance, the lesion-wise sensitivity, precision, and average false positive rate (FPR) per patient \modified{were} used for the evaluation of BM identification accuracy. 
In this work, any segmentation demonstrating an overlap of at least one voxel with the reference label segmentation was considered a true positive (i.e., no overlap cutoff), since tiny BM contain only a few voxels and some comprise just a single voxel.
Because of the trade-off between sensitivity and precision, F$_\beta$ score is commonly used in many biomedical detection and segmentation tasks:
\begin{equation}
\textrm{F}_\beta = (1 + \beta^2)\cdot \frac{\textrm{sensitivity} \cdot \textrm{precision}}{\beta^2\cdot \textrm{precision} + \textrm{sensitivity}},
\end{equation}
where a larger $\beta$ values sensitivity more than precision. In this work, F1 and F2 are used.

In clinical practice, true positive metastases are confirmed first and subsequently their contouring masks will be refined for treatment planning (i.e., \textbf{detection first and contouring later}) in a manual manner. Therefore, in this work, the contouring accuracy is evaluated on a per-lesion level for true positive metastases only. \modified{In addition to the conventional Dice score and Hausdorff distance, the surface Dice (sDice) score \cite{nikolov2021clinically} with a tolerance of 1\,mm and 95\% Hausdorff distance (HD95) are applied to compensate annotation errors.} The definition of sDice is provided in \cite{nikolov2021clinically}, and HD95 is calculated as the 95th percentile of the distances between boundary points of two segmentation masks. All the experiments were repeated 5 times and the unpaired t-test was used for assessing statistical significance.

\begin{table*}[t]
\caption{{Autosegmentation performance on different datasets with single-center or mixed training (average from 5 repeats, lesion-wise, HD95 unit: mm).}}
\label{Tab:PerformanceTable}
\centering
\begin{tabular}{|l|l|lcccccc|}
\noalign{\hrule height 1.5pt}
Training & Test Data  & Sensitivity & Precision & FPR&F1 & F2  &sDice &HD95\\
%\hline
\noalign{\hrule height 1.5pt}
\multirow{7}{*}{{Single-center}}&UKER & {0.854} &{0.900} &0.25 &{0.876} &{0.863}  &{0.790}&4.21\\
\cline{2-9}
&\multirow{2}{*}{USZ}  & \multirow{2}{*}{{0.831}} &{0.776} &0.99 &{0.802} &{0.819} &\multirow{2}{*}{{0.727}} &\multirow{2}{*}{8.43} \\
&  & & ({0.367}) &(6.06) &({0.506}) & ({0.659}) & &\\
\cline{2-9}
&Stanford  &{0.657} &{0.723} &4.15 &{0.688} &{0.669}  &{0.671}&3.77\\
\cline{2-9}
&UCSF &{0.847} & {0.704} &2.94 &{0.769} &{0.814} & {0.759}&3.67\\
\cline{2-9}
&NYU & {0.714} &{0.557} & 4.11 & {0.625} & {0.675}  & {0.600}& 8.61\\
\cline{2-9}
&BraTS &{0.769} & {0.575} &2.17 & {0.658} &{0.720} & {0.595}&6.79 \\
\noalign{\hrule height 1.5pt}
&UKER &{0.906} &{0.814} & 0.55 &{0.857} &\textbf{{0.886}} & {0.880} &2.75\\
\cline{2-2}\cline{3-9}
&USZ & {0.885} & {0.651} & 1.95 & {0.750} & {0.825}  & {0.775}&8.76\\
\cline{2-2}\cline{3-9}
Mixed (five centers&Stanford & {0.781} & {0.836} & 2.54 & {0.807} & \textbf{{0.791}}  &{0.839} &2.27\\
\cline{2-2}\cline{3-9}
without USZ)&UCSF & {0.855} &{0.705} &2.98 &{0.772} &{0.820} &{0.839} &2.19\\
\cline{2-2}\cline{3-9}
&NYU & {0.764} & {0.649} & 2.98 & {0.702} & \textbf{{0.738}}  & {0.767} &5.55\\
\cline{2-2}\cline{3-9}
&BraTS &{0.867} &{0.445} & 4.64 & {0.587} &{0.728}  &{0.689} &5.83\\
\noalign{\hrule height 1.5pt}
\end{tabular}

      \small{{\raggedright Note: The values in brackets for USZ are without binary brain masks (brain masks can remove false positive metastases outside brain regions). Others have the same values for both with and without brain masks. The bold values highlight the centers, which have notable benefit in BM detection from mixed training compared with {single-center} training. \par}}
\end{table*}

\section{Results}

\begin{table*}[t]
\caption{{Autosegmentation performance for bilateral collaboration using different training methods (average from 5 repeats, lesion-wise, HD95 unit: mm).}}
\label{Tab:PerformanceTableLWF}
\centering
\begin{tabular}{|l|l|l|ccccccc|}
\noalign{\hrule height 1.5pt}
Training & Test Data &Model & Sensitivity & Precision & FPR&F1 & F2  &sDice &HD95\\
%\hline
\noalign{\hrule height 1.5pt}
 & & \multirow{2}{*}{USZ\textsubscript{{single-center}}} &\multirow{2}{*}{{0.839}} & {0.584} &{1.81} & {0.688} &{0.771}   &\multirow{2}{*}{{0.755}} &\multirow{2}{*}{5.36}\\
& Combined & & &({0.262}) & ({7.24}) & ({0.399}) &({0.581})  & &\\
\cline{3-10}
UKER+USZ&UKER &\multirow{2}{*}{TL\textsubscript{UKER$\Rightarrow$USZ}} & \multirow{2}{*}{{\textbf{0.905}}} &{0.418} &{3.89}  &{0.570}  &{0.732} &\multirow{2}{*}{\textbf{{0.821}}}&\multirow{2}{*}{\textbf{{4.63}}}\\
collaboration&+USZ & & &({0.356}) & ({5.26}) & ({0.506}) & ({0.685}) &  &\\
\cline{3-10}
& & LWF\textsubscript{UKER$\Rightarrow$USZ} & {0.864} &{\textbf{0.815}}  &0.59  &\textbf{{0.839}}  &\textbf{{0.854}}  &{0.737} &7.01 \\
\noalign{\hrule height 1.5pt}
& & Stanford\textsubscript{{single-center}} &{0.686} & {0.591} &3.29 & {0.630} & {0.661}  &{0.592} &7.79\\
\cline{3-10}
UKER+&Combined &TL\textsubscript{UKER$\Rightarrow$Stanford} & {\textbf{0.854}} &{0.455} & 6.66 & {0.593} & {0.726}  &\textbf{{0.823}} &\textbf{2.77}\\
\cline{3-10}
Stanford&UKER+ & LWF\textsubscript{UKER$\Rightarrow$Stanford} & {0.763} & {\textbf{0.811}} & 1.16 & \textbf{{0.786}} &\textbf{{0.772}} & {0.779} &3.51\\
\cline{3-10}
collaboration&Stanford & LWF\textsubscript{UKER$\Rightarrow$Stanford, CWT} & {0.796} & {0.708} & 2.15 & \textbf{{0.749}} &\textbf{{0.776}} & {0.798} &3.13\\
\cline{3-10}
& & Mixed\textsubscript{UKER$+$Stanford} & {0.783} & {0.832} & 1.03 & {0.807} &{0.792} & {0.827} &2.85\\
\noalign{\hrule height 1.5pt}
\end{tabular}

      \small{{\raggedright Note: The values in brackets are without binary brain masks (brain masks can remove false positive metastases outside brain regions). Others have the same values for both with and without brain masks. TL: transfer learning (TL) without LWF; LWF: transfer learning with LWF; CWT: cyclic weight transfer. The bold values highlight the best detection or contouring performances. {USZ\textsubscript{single-center} and Stanford\textsubscript{single-center} were solely trained on USZ and Stanford data respectively and are provided as a reference.}\par}}

\end{table*}

{\textbf{Data heterogeneity:}} In addition to Tab.\,\ref{Tab:Datasets}, axial and coronal image examples from different centers are displayed in Fig.\,\ref{Fig:DataExamples}, where data heterogeneity is clearly visualized:

a) BM density: The UKER dataset has the lowest density, whereas Stanford and UCSF have a high density.

b) BM spatial distribution: The USZ dataset includes metastases of leptomeningeal origin near cortical surfaces and meninges, while BM in other datasets are mainly located in the brain parenchyma.

c) Image resolution: Many cases in the NYU and BraTS datasets have low resolution along the transversal dimension (due to the use of older 2D T1 spin echo sequences), which increases the difficulty for human labeling as well as automatic detection.

d) Image contrast: The UCSF datasets contains much more cases where blood vessels and meninges are highly enhanced (e.g. related to higher contrast agent dose or lower time interval between contrast administration and sequence acquisition in the imaging protocols), which has the risk to increase false positive detection rate, e.g., the false positive structure in Fig.\,\ref{Fig:DataExamples}(d).

\bigskip
{\textbf{Single-center training:}} \modified{The single-center training performances of UKER and UCSF models with respect to training data amount are displayed in Fig.\,\ref{Fig:PerformanceOverDataAmount}(a) and (b), respectively. They demonstrate the advantage of F2 over F1 as a BM detection metric.} The detailed performance of all the single-center-training models is displayed in the top part of Tab.\,\ref{Tab:PerformanceTable}. 
The UKER model achieves a relatively high lesion-wise sensitivity and precision, potentially due to the high image quality, low metastasis density, and a large number of training volumes. The UCSF model achieves a similar sensitivity, but with a slightly lower precision. The Stanford model achieves a low sensitivity, due to the small number of training patients. In contrast, the NYU and BraTS models achieved low precision values on their respective test data, which could be ascribed to their low axial slice resolution.
As the USZ dataset includes metastases of leptomeningeal origin near cortical surfaces and meninges, the trained model predicted many false positive metastases outside the brain region, e.g., Fig.\,\ref{Fig:BMIdentificationExamples}(b) and Fig.\,\ref{Fig:BMIdentificationExamples}(e). This behavior is only observed for models trained from the USZ dataset, but serves as a good indicator of model performance and knowledge transferability. 
Such false positive metastases can be simply removed with binary brain masks, and the precision is improved from 0.367 to 0.776.

\bigskip
{\textbf{Mixed training:}} \modified{The detailed performance of mixed multicenter training (USZ data excluded due to data privacy) is displayed} in the bottom part of Tab\,\ref{Tab:PerformanceTable}. {Compared to single-center training, mixed training demonstrates improved detection sensitivity across all centers, with statistically significant (p $\leq$ \modified{0.01}) gains at all centers except UCSF. The F2 scores of Stanford (p $\leq$ \modified{0.01}), NYU (p $\leq$ \modified{0.01}), and UKER (p \modified{$\leq$ 0.02}) have a significant improvement, as displayed in Fig.\,\ref{Fig:PerformanceOverDataAmount}(c).} 
However, USZ, UCSF and BraTS with mixed training have similar F2 scores to those with single-center training, respectively, as displayed in Fig.\,\ref{Fig:PerformanceOverDataAmount}(c). 
 \textbf{This observation indicates that more training data from other centers does not necessarily improve BM detection performance, which can be attributed to the data heterogeneity across centers.}
 Nevertheless, the mixed-training model achieved \modified{significantly better (p $\leq$ 0.01 except USZ)} contouring performance  for true positive metastases than single-center-training models, as displayed in Fig.\,\ref{Fig:PerformanceOverDataAmount}(d).

\bigskip
{\textbf{UKER model shared to USZ:}} The results of the bilateral collaboration between UKER and USZ are reported in Tab.\,\ref{Tab:PerformanceTableLWF} (overall performance on both test datasets) as well as supplementary Tab. 2 (performance on each individual test dataset):

{a) Single-center training:} When the UKER pretrained model was directly applied to USZ test data without further fine-tuning, the model achieved decent sensitivity and precision (see Supplementary Tab. 2). Even when no brain masks were used, no false positive metastases were predicted outside the brain region.  However, when the USZ pretrained model was directly applied to UKER test data, the model obtained a decent sensitivity but a very low precision. {\textbf{This indicates the good generalizability of the UKER model to the USZ data, whereas the reverse scenario--applying the model from USZ to UKER data--does not possess comparable generalizability.}}

{b) Naive transfer learning (TL):} When the UKER model was fine-tuned without LWF on USZ training data, such a naive TL model achieved high sensitivity (0.915) but low precision (0.362) for the UKER test data. The TL model predicted false positive metastases near the brain boundary for both USZ and UKER data (some examples in Fig.\,\ref{Fig:BMIdentificationExamples}(c)-(h)). {\textbf{These results reveal that the fine-tuned model overfits to the USZ data and forgets the knowledge learned from the UKER data.}} 

{c) LWF:} The LWF model achieved decent sensitivity (0.822) and precision (0.730) without brain mask for the USZ test data. No false positive metastases were predicted outside the brain region, {\textbf{indicating the benefit of LWF in the target center}. The LWF model achieved high sensitivity (0.874) and high precision (0.855) on the UKER test data} (one example in Fig. \ref{Fig:BMIdentificationExamples}(d)), {\textbf{demonstrating the learned knowledge from the UKER training dataset was preserved}}. 
 
d) Overall comparison: The USZ single-center-training model and the TL model achieved low F2 scores with brain masks on combined test data (Tab.\,\ref{Tab:PerformanceTableLWF}), respectively. In contrast, \textbf{the LWF model achieved the best F2 score of 0.854, which is significantly better than USZ single-center training and TL (\modified{p $\leq$ 0.01, see Fig.\,\ref{Fig:PerformanceOverDataAmount}(e)}). This highlights the benefit of LWF in privacy-preserving bilateral collaboration for BM detection.} Although TL is not optimal for BM detection, it is beneficial to improve BM contouring accuracy of true positive metastases (sDice significant with \modified{p $\leq$ 0.01, see Fig.\,\ref{Fig:PerformanceOverDataAmount}(e)}).

\bigskip
{\textbf{UKER model shared to Stanford:}} The main differences between the UKER and Stanford datasets are the BM density, BM size, and the number of training patients. The Stanford single-center-training model got low sensitivity and low precision on the combined test data, because of the low number of training data. The TL model has a significantly large improvement in detection sensitivity (p $\leq$ \modified{0.01}) and contouring accuracy (p $\leq$ \modified{0.01}). But it has a very low detection precision (p \modified{$\leq$ 0.03}). Likely the TL model tends to overestimate the presence of BM in the UKER test data due to the high BM density in the Stanford training data. The LWF model achieved both good sensitivity and precision, and hence has a significantly better F2 score of 0.772 than single-center training (p $\leq$ \modified{0.01}) and TL (p $\leq$ \modified{0.01}), which is close to that of mixed training (F2 = 0.792). Compared with single weight transfer (SWT), CWT with 4 iterations using LWF further improves the segmentation accuracy (sDice from 0.779 to 0.798, HD95 from 3.51\,mm to 3.13\,mm), but no statistical significance was found.

\section{Discussion}

The results on the UKER-USZ and UKER-Stanford collaborations show that TL can improve the sensitivity of the model as well as the segmentation accuracy of true positive metastases compared with single-center training, but the TL model tends to overfit to the target center data and detects more false positive BM. Further comparing LWF with TL, LWF can achieve relatively high sensitivity as well as high precision. This indicates that LWF is a promising approach for peer-to-peer federated learning and multicenter collaboration on BM autosegmentation without sharing raw data.

Several multicenter studies on deep learning for BM autosegmentation have been published previously  \cite{yin2022development,buchner2023development,qu2023construction,ottesen20232,liew2023gradual,bouget2022preoperative}.  In contrast to the AURORA multicenter study with a median BM volume size of 7.3\,cm$^3$, the datasets in our study feature significantly smaller median volume sizes, all less than 0.5\,cm$^3$. The principal novelty of our work lies in the introduction of a privacy-preserving method, i.e. LWF, which enables the training of a model across multiple centers without the need to share raw data. This contrasts with other studies, which typically rely on pooled multicenter data for model training. This approach addresses significant privacy concerns and enhances the feasibility of multicenter collaboration on BM autosegmentation.

The {single-center} training shows that Stanford, NYU and BraTS had low F2 scores. With mixed training, the F2 scores of NYU and BraTS stayed low despite of improvement. This is very likely because of their distinct image features from other datasets, in particular the low axial slice resolution. {\textbf{This implies that deep learning models for BM identification also require high image quality similar to human experts.}} For human experts, $\leq $1\,mm isotropic resolution on 3D scans has become a consensus of image resolution requirement among radiation oncology communities \cite{putz2024quality}. Human experts have also reached consensus on optimal sequence selection, contrast agent uptake, distortion correction, and motion control \cite{putz2024quality}. Satisfying such requirements are believed to improve deep learning model performance for BM identification as well.

In general medical tasks, F1 is the most widely used F$_\beta$ score. For BM detection, the first primal clinical requirement is high sensitivity in detecting BM, regardless of the exact 3D extent of each metastasis. False positives can be removed afterwards by human expert review. In practice, achieving high sensitivity is more challenging than achieving high precision. For example, a network can achieve high precision by detecting a few large, high-contrast BM, but it is very challenging to detect all the metastases (large and small, high and low contrast). \textbf{Therefore, F2, which prioritizes sensitivity, is more appropriate than F1 for BM detection in clinical application. Fig.\,\ref{Fig:PerformanceOverDataAmount}(a)-(b) and the average detection performance of Tab. 2 in \cite{rudie2023university} demonstrate the advantage of F2 over F1 with increasing training data amount.}  Note that the benefit of LWF over TL and single-center training is more prominent in F1 than F2, as displayed in Fig.\,\ref{Fig:PerformanceOverDataAmount} and Tab.\,\ref{Tab:PerformanceTableLWF}.

{This work has several limitations that should be noted: a) The training of 3D models requires significant computational resources, which limited us to only five repetitions for each experimental setting. Ideally, a greater number of repetitions would be conducted to enhance the statistical robustness of the results. b) Although six datasets from different centers were available for this study, this offers numerous potential combinations for multicenter collaboration. However, we only investigated two bilateral collaboration scenarios to demonstrate the benefits of LWF. We plan to explore additional combinations and collaboration models in our future work to further validate and expand on these initial findings.}

In addition, future studies will aim to include a wider variety of datasets from different hospitals and institutions across Europe. This would test the model's robustness and ability to generalize across diverse patient demographics and equipment variations. The ultimate goal is to seamlessly integrate the method in this work into existing clinical workflows. Achieving this integration necessitates comprehensive AI-specific quality assurance (QA) measures \cite{claessens2022quality} to ensure the reliability, accuracy, and safety of the model. Prior to clinical deployment, the model must undergo stringent validation and secure approval from relevant health authorities, such as the Food and Drug Administration (FDA) or European Medicines Agency (EMA). Significant additional research will be necessary to bridge the gap between the developmental phase and practical clinical applications, ensuring that all scientific, regulatory, and operational benchmarks are met.

\section{Conclusion}
The heterogeneity of data across multiple centers poses a significant challenge to the generalizability of deep learning models from one center to another. This is particularly evident in the development of deep learning models for BM identification, which, akin to human experts, necessitate high-quality imaging data for both training and testing.  
TL and LWF emerge as valuable strategies for fostering privacy-preserving collaborations in the advancement of BM autosegmentation models. While TL demonstrates notable strengths in achieving high detection sensitivity and contouring accuracy, it does so at the expense of precision. Conversely, LWF presents a commendable equilibrium between sensitivity and precision, offering a more balanced approach to multicenter model development.

\biboptions{sort&compress}
%\bibliography{refs}

\begin{thebibliography}{10}
\expandafter\ifx\csname url\endcsname\relax
  \def\url#1{\texttt{#1}}\fi
\expandafter\ifx\csname urlprefix\endcsname\relax\def\urlprefix{URL }\fi
\expandafter\ifx\csname href\endcsname\relax
  \def\href#1#2{#2} \def\path#1{#1}\fi

\bibitem{tabouret2012recent}
E.~Tabouret, O.~Chinot, P.~Metellus, A.~Tallet, P.~Viens, A.~Goncalves, Recent
  trends in epidemiology of brain metastases: an overview, Anticancer Res.
  32~(11) (2012) 4655--4662.

\bibitem{rogers2022stereotactic}
S.~Rogers, B.~Baumert, O.~Blanck, D.~B{\"o}hmer, J.~Bostr{\"o}m,
  R.~Engenhart-Cabillic, E.~Ermis, S.~Exner, M.~Guckenberger, D.~Habermehl,
  et~al., Stereotactic radiosurgery and radiotherapy for resected brain
  metastases: current pattern of care in the radiosurgery and stereotactic
  radiotherapy working group of the german association for radiation oncology
  {(DEGRO)}, Strahlentherapie und Onkologie 198~(10) (2022) 919--925.

\bibitem{welzel2022stereotactic}
T.~Welzel, R.~A. El~Shafie, B.~v.~Nettelbladt, D.~Bernhardt, S.~Rieken,
  J.~Debus, Stereotactic radiotherapy of brain metastases: clinical impact of
  three-dimensional {SPACE} imaging for {3T-MRI-based} treatment planning,
  Strahlentherapie und Onkologie 198~(10) (2022) 926--933.

\bibitem{xue2020deep}
J.~Xue, B.~Wang, Y.~Ming, X.~Liu, Z.~Jiang, C.~Wang, X.~Liu, L.~Chen, J.~Qu,
  S.~Xu, et~al., Deep learning--based detection and segmentation-assisted
  management of brain metastases, Neuro Oncol. 22~(4) (2020) 505--514.

\bibitem{kocher2020applications}
M.~Kocher, M.~I. Ruge, N.~Galldiks, P.~Lohmann, Applications of radiomics and
  machine learning for radiotherapy of malignant brain tumors, Strahlenther.
  Onkol. 196 (2020) 856--867.

\bibitem{wang2023brain}
T.-W. Wang, M.-S. Hsu, W.-K. Lee, H.-C. Pan, H.-C. Yang, C.-C. Lee, Y.-T. Wu,
  Brain metastasis tumor segmentation and detection using deep learning
  algorithms: a systematic review and meta-analysis, Radiotherapy and Oncology
  (2023) 110007.

\bibitem{ozkara2023deep}
B.~B. Ozkara, M.~M. Chen, C.~Federau, M.~Karabacak, T.~M. Briere, J.~Li,
  M.~Wintermark, Deep learning for detecting brain metastases on {MRI}: a
  systematic review and meta-analysis, Cancers 15~(2) (2023) 334.

\bibitem{buchner2023development}
J.~A. Buchner, F.~Kofler, L.~Etzel, M.~Mayinger, S.~M. Christ, T.~B. Brunner,
  A.~Wittig, B.~Menze, C.~Zimmer, B.~Meyer, et~al., Development and external
  validation of an {MRI-based} neural network for brain metastasis segmentation
  in the {AURORA} multicenter study, Radiotherapy and Oncology 178 (2023)
  109425.

\bibitem{qu2023construction}
J.~Qu, W.~Zhang, X.~Shu, Y.~Wang, L.~Wang, M.~Xu, L.~Yao, N.~Hu, B.~Tang,
  L.~Zhang, et~al., Construction and evaluation of a gated high-resolution
  neural network for automatic brain metastasis detection and segmentation,
  European Radiology (2023) 1--11.

\bibitem{yin2022development}
S.~Yin, X.~Luo, Y.~Yang, Y.~Shao, L.~Ma, C.~Lin, Q.~Yang, D.~Wang, Y.~Luo,
  Z.~Mai, et~al., Development and validation of a deep-learning model for
  detecting brain metastases on {3D} post-contrast {MRI}: a multi-center
  multi-reader evaluation study, Neuro-oncology 24~(9) (2022) 1559--1570.

\bibitem{ottesen20232}
J.~A. Ottesen, D.~Yi, E.~Tong, M.~Iv, A.~Latysheva, C.~Saxhaug, K.~D. Jacobsen,
  {\AA}.~Helland, K.~E. Emblem, D.~L. Rubin, et~al., 2.5d and 3d segmentation
  of brain metastases with deep learning on multinational {MRI} data, Frontiers
  in Neuroinformatics 16 (2023) 1056068.

\bibitem{liew2023gradual}
A.~Liew, C.~C. Lee, V.~Subramaniam, B.~L. Lan, M.~Tan, Gradual self-training
  via confidence and volume based domain adaptation for multi dataset deep
  learning-based brain metastases detection using nonlocal networks on {MRI}
  images, Journal of Magnetic Resonance Imaging 57~(6) (2023) 1728--1740.

\bibitem{bouget2022preoperative}
D.~Bouget, A.~Pedersen, A.~S. Jakola, V.~Kavouridis, K.~E. Emblem, R.~S.
  Eijgelaar, I.~Kommers, H.~Ardon, F.~Barkhof, L.~Bello, et~al., Preoperative
  brain tumor imaging: Models and software for segmentation and standardized
  reporting, Frontiers in neurology 13 (2022) 932219.

\bibitem{beckers2021eu}
R.~Beckers, Z.~Kwade, F.~Zanca, The {EU} medical device regulation:
  Implications for artificial intelligence-based medical device software in
  medical physics, Phys. Med. 83 (2021) 1--8.

\bibitem{xu2021federated}
J.~Xu, B.~S. Glicksberg, C.~Su, P.~Walker, J.~Bian, F.~Wang, Federated learning
  for healthcare informatics, J. Healthc. Inform. Res. 5~(1) (2021) 1--19.

\bibitem{sheller2020federated}
M.~J. Sheller, B.~Edwards, G.~A. Reina, J.~Martin, S.~Pati, A.~Kotrotsou,
  M.~Milchenko, W.~Xu, D.~Marcus, R.~R. Colen, et~al., Federated learning in
  medicine: facilitating multi-institutional collaborations without sharing
  patient data, Scientific reports 10~(1) (2020) 12598.

\bibitem{delange2021continual}
M.~Delange, R.~Aljundi, M.~Masana, S.~Parisot, X.~Jia, A.~Leonardis,
  G.~Slabaugh, T.~Tuytelaars, A continual learning survey: Defying forgetting
  in classification tasks, IEEE Trans. Pattern Anal. Mach. Intell. 1 (2021)
  1--20.

\bibitem{huang2023survey}
Y.~Huang, C.~Bert, A.~Gomaa, R.~Fietkau, A.~Maier, F.~Putz, An experimental
  survey of incremental transfer learning for multicenter collaboration, IEEE
  Access (2024) 1--16(in press).

\bibitem{li2017learning}
Z.~Li, D.~Hoiem, Learning without forgetting, IEEE Trans. Pattern Anal. Mach.
  Intell. 40~(12) (2017) 2935--2947.

\bibitem{grovik2020deep}
E.~Gr{\o}vik, D.~Yi, M.~Iv, E.~Tong, D.~Rubin, G.~Zaharchuk, Deep learning
  enables automatic detection and segmentation of brain metastases on
  multisequence {MRI}, J. Magn. Reson. Imaging 51~(1) (2020) 175--182.

\bibitem{rudie2023university}
J.~D. Rudie, R.~S. D.~A. Weiss, P.~Nedelec, E.~Calabrese, J.~B. Colby,
  B.~Laguna, J.~Mongan, S.~Braunstein, C.~P. Hess, A.~M. Rauschecker, et~al.,
  {The University of California San Francisco}, brain metastases stereotactic
  radiosurgery {(UCSF-BMSR)} {MRI} dataset, arXiv preprint arXiv:2304.07248
  (2023) 1--18.

\bibitem{moawad2023brain}
A.~W. Moawad, A.~Janas, U.~Baid, D.~Ramakrishnan, L.~Jekel, K.~Krantchev,
  H.~Moy, R.~Saluja, K.~Osenberg, K.~Wilms, et~al., The brain tumor
  segmentation ({BraTS-METs}) challenge 2023: Brain metastasis segmentation on
  pre-treatment mri, ArXiv (2023) 1--16.

\bibitem{putz2024quality}
F.~Putz, M.~Bock, D.~Schmitt, C.~Bert, O.~Blanck, M.~I. Ruge, E.~Hattingen,
  C.~P. Karger, R.~Fietkau, J.~Grigo, et~al., Quality requirements for {MRI}
  simulation in cranial stereotactic radiotherapy: a guideline from the german
  taskforce “imaging in stereotactic radiotherapy”, Strahlentherapie und
  Onkologie (2024) 1--18.

\bibitem{kaufmann2020consensus}
T.~J. Kaufmann, M.~Smits, J.~Boxerman, R.~Huang, D.~P. Barboriak, M.~Weller,
  C.~Chung, C.~Tsien, P.~D. Brown, L.~Shankar, et~al., Consensus
  recommendations for a standardized brain tumor imaging protocol for clinical
  trials in brain metastases, Neuro-oncology 22~(6) (2020) 757--772.

\bibitem{buchner2023identifying}
J.~A. Buchner, J.~C. Peeken, L.~Etzel, I.~Ezhov, M.~Mayinger, S.~M. Christ,
  T.~B. Brunner, A.~Wittig, B.~H. Menze, C.~Zimmer, et~al., Identifying core
  {MRI} sequences for reliable automatic brain metastasis segmentation,
  Radiotherapy and Oncology 188 (2023) 109901.

\bibitem{putz2020fsrt}
F.~Putz, T.~Weissmann, D.~Oft, M.~A. Schmidt, J.~Roesch, H.~Siavooshhaghighi,
  I.~Filimonova, C.~Schmitter, V.~Mengling, C.~Bert, et~al., {FSRT} vs. {SRS}
  in brain metastases—differences in local control and radiation necrosis—a
  volumetric study, Front. Oncol. 10 (2020).

\bibitem{kamnitsas2017efficient}
K.~Kamnitsas, C.~Ledig, V.~F. Newcombe, J.~P. Simpson, A.~D. Kane, D.~K. Menon,
  D.~Rueckert, B.~Glocker, Efficient multi-scale {3D} {CNN} with fully
  connected {CRF} for accurate brain lesion segmentation, Med. Image Anal. 36
  (2017) 61--78.

\bibitem{liu2017deep}
Y.~Liu, S.~Stojadinovic, B.~Hrycushko, Z.~Wardak, S.~Lau, W.~Lu, Y.~Yan, S.~B.
  Jiang, X.~Zhen, R.~Timmerman, et~al., A deep convolutional neural
  network-based automatic delineation strategy for multiple brain metastases
  stereotactic radiosurgery, PloS One 12~(10) (2017) e0185844.

\bibitem{lu2019automated}
S.~Lu, S.~Hu, W.~Weng, Y.~Chen, J.~Lu, F.~Xiao, F.~Hsu, Automated detection and
  segmentation of brain metastases in stereotactic radiosurgery using
  three-dimensional deep neural networks, Int. J. Radiat. Oncol. Biol. Phys.
  105~(1) (2019) S69--S70.

\bibitem{charron2018automatic}
O.~Charron, A.~Lallement, D.~Jarnet, V.~Noblet, J.-B. Clavier, P.~Meyer,
  Automatic detection and segmentation of brain metastases on multimodal {MR}
  images with a deep convolutional neural network, Comput. Biol. Med. 95 (2018)
  43--54.

\bibitem{hu2019multimodal}
S.-Y. Hu, W.-H. Weng, S.-L. Lu, Y.-H. Cheng, F.~Xiao, F.-M. Hsu, J.-T. Lu,
  Multimodal volume-aware detection and segmentation for brain metastases
  radiosurgery, in: Proc. AIRT, Springer, 2019, pp. 61--69.

\bibitem{huang2022deep}
Y.~Huang, C.~Bert, P.~Sommer, B.~Frey, U.~Gaipl, L.~V. Distel, T.~Weissmann,
  M.~Uder, M.~A. Schmidt, A.~D{\"o}rfler, et~al., Deep learning for brain
  metastasis detection and segmentation in longitudinal mri data, Medical
  Physics 49~(9) (2022) 5773--5786.

\bibitem{chang2018distributed}
K.~Chang, N.~Balachandar, C.~Lam, D.~Yi, J.~Brown, A.~Beers, B.~Rosen, D.~L.
  Rubin, J.~Kalpathy-Cramer, Distributed deep learning networks among
  institutions for medical imaging, Journal of the American Medical Informatics
  Association 25~(8) (2018) 945--954.

\bibitem{hinton2015distilling}
G.~Hinton, O.~Vinyals, J.~Dean, Distilling the knowledge in a neural network,
  in: NIPS Deep Learning and Representation Learning Workshop, 2015, pp. 1--9.

\bibitem{nikolov2021clinically}
S.~Nikolov, S.~Blackwell, A.~Zverovitch, R.~Mendes, M.~Livne, J.~De~Fauw,
  Y.~Patel, C.~Meyer, H.~Askham, B.~Romera-Paredes, et~al., Clinically
  applicable segmentation of head and neck anatomy for radiotherapy: deep
  learning algorithm development and validation study, Journal of medical
  Internet research 23~(7) (2021) e26151.

\bibitem{claessens2022quality}
M.~Claessens, C.~S. Oria, C.~L. Brouwer, B.~P. Ziemer, J.~E. Scholey, H.~Lin,
  A.~Witztum, O.~Morin, I.~El~Naqa, W.~Van~Elmpt, et~al., Quality assurance for
  ai-based applications in radiation therapy, in: Seminars in radiation
  oncology, Vol.~32, Elsevier, 2022, pp. 421--431.

\end{thebibliography}

\begin{figure*}
\centering
\includegraphics[width=0.72\linewidth]{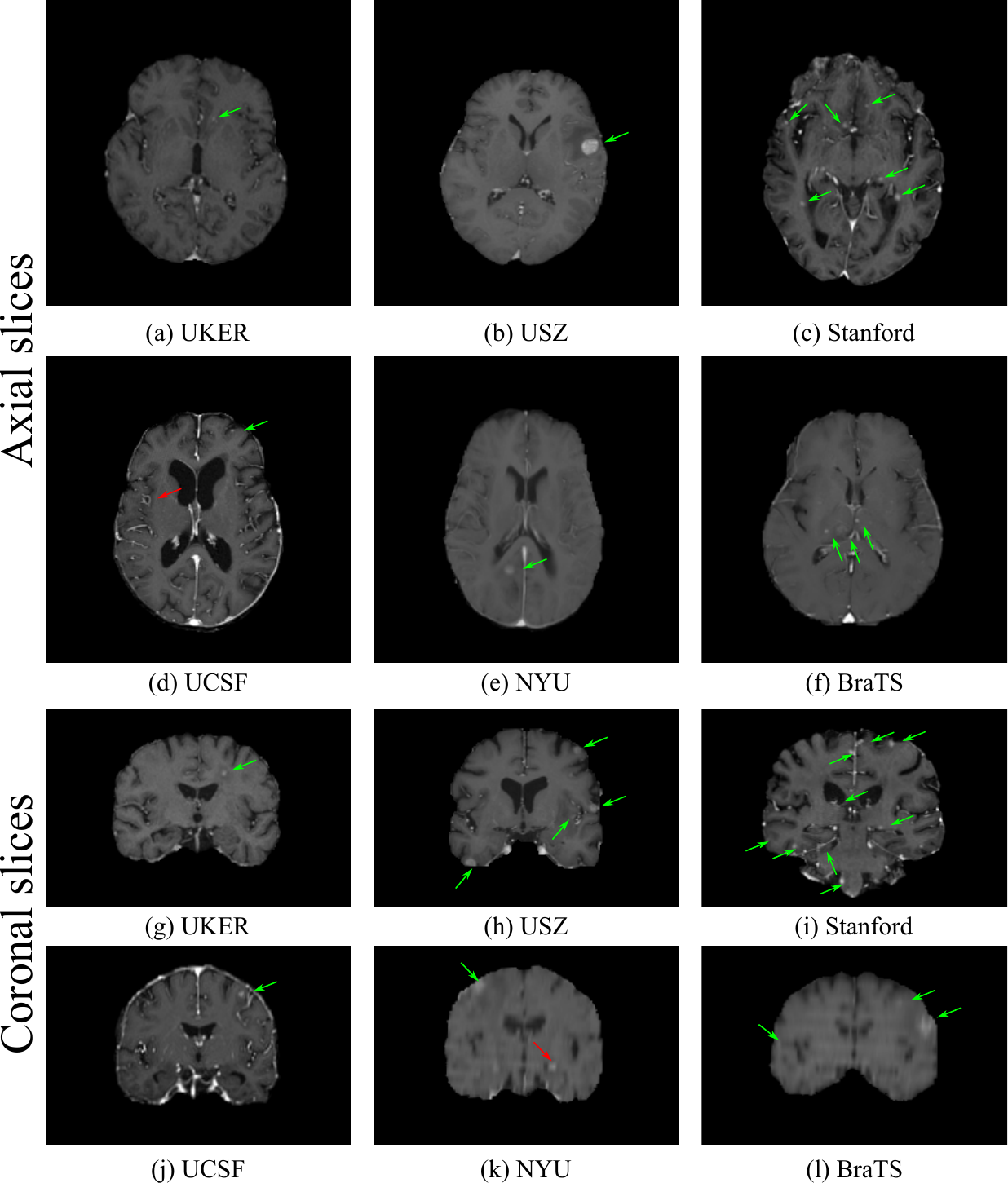}

\caption{Exemplary images from different datasets (top: axial slices; bottom: coronal slices). The axial and coronal slices of the same center are not from the same patients. The {green} arrows indicate true positive brain metastases, and the {red} arrows indicate suspicious spots which are true negative.}
\label{Fig:DataExamples}
\end{figure*}

\begin{figure*}
\centering
\includegraphics[width=0.76\linewidth]{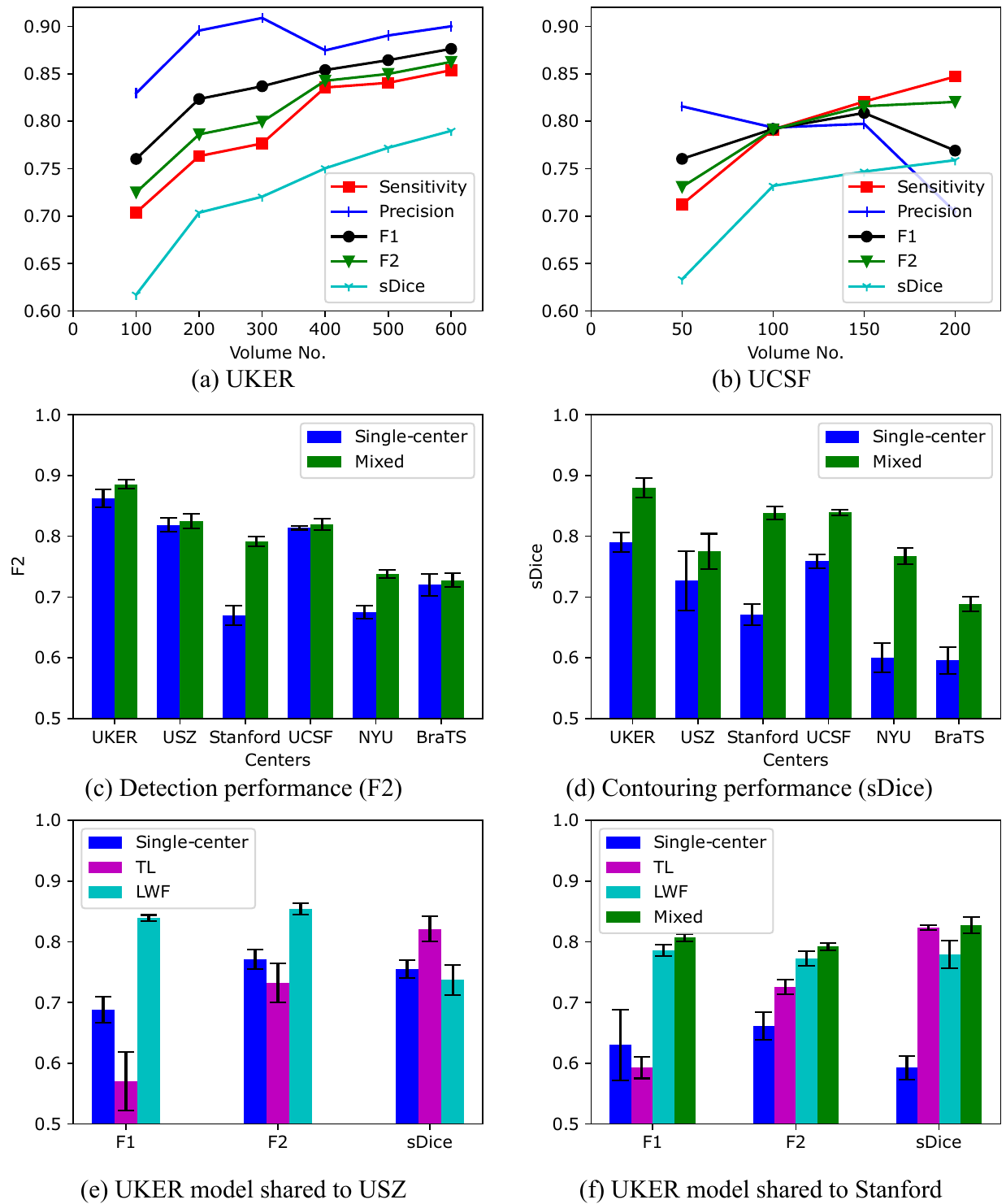}
\caption{The BM detection and segmentation performances of different models. (a) and (b) are the performances of UKER and UCSF {single-center-training} models with respect to training data amount (number of volumes/patients), respectively. (c) and (d) display the BM detection performance and segmentation performance respectively with {single-center} training and mixed training. (e) and (f) display BM detection and segmentation performances of different methods. \modified{The error bars in (c)-(d) indicate the standard deviations.} }
\label{Fig:PerformanceOverDataAmount}
\end{figure*}

\begin{figure*}
\includegraphics[width=\linewidth]{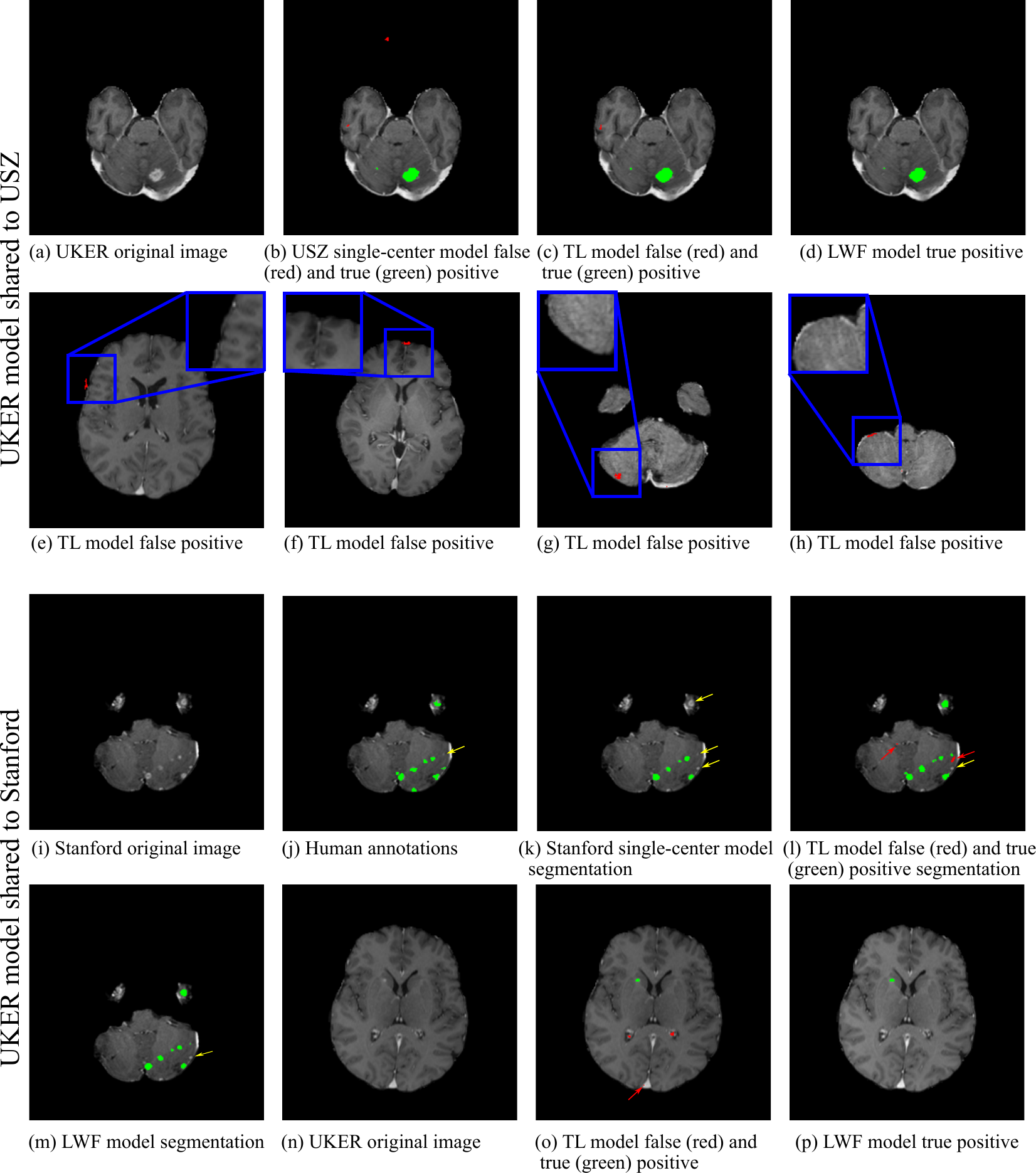}

\caption{The BM autosegmentation examples of different models. The top rows displays the results of different models on one UKER exemplary image when the UKER model was shared to USZ, while the second rows displays other representative false positive examples of the TL\textsubscript{UKER$\Rightarrow$USZ} model. The bottom two rows display the results of different models when the UKER model was shared to Stanford. {Red} areas are false positive segmentations, {green} areas are true positive segmentations, and the yellow arrows indicate false negative metastases. {Subfigure (j) is an example of human annotation errors in the Stanford dataset, where the tiny metastasis indicated by the yellow arrow was not labeled, and the annotation mask of the top metastasis is not accurate.}}
\label{Fig:BMIdentificationExamples}
\end{figure*}

\end{document}